\documentclass[onecolumn,preprintnumbers,amsmath,amssymb,eqsecnum,11pt]{revtex4}

\usepackage{epsf}
\usepackage{graphicx}  
\usepackage{dcolumn}   
\usepackage{bm}        

\begin{document}


\title{Unified First Law and Thermodynamics of Apparent Horizon
in FRW Universe}

 \author{Rong-Gen Cai\footnote{e-mail address:
cairg@itp.ac.cn}}

\address{
  Institute of Theoretical Physics, Chinese
Academy of Sciences, P.O. Box 2735, Beijing 100080, China}

\author{Li-Ming Cao\footnote{e-mail address:
caolm@itp.ac.cn}}
\address{Institute of Theoretical Physics, Chinese
Academy of Sciences,
 P.O. Box 2735, Beijing 100080, China\\
  Graduate School of the Chinese Academy of Sciences, Beijing 100039, China}

\begin{abstract}
In this paper we revisit the relation between the Friedmann
equations and the first law of thermodynamics. We find that the
unified first law firstly proposed by Hayward to treat the
``outer"trapping horizon of dynamical black hole can be used to the
apparent horizon (a kind of ``inner" trapping horizon in the context
of the FRW cosmology) of the FRW universe. We  discuss three kinds
of gravity theorties: Einstein theory, Lovelock thoery and
scalar-tensor theory. In Einstein theory, the first law of
thermodynamics is always satisfied on the apparent horizon. In
Lovelock theory, treating the higher derivative terms as an
effective energy-momentum tensor, we find that this method can give
the same entropy formula for the apparent horizon as that of black
hole horizon. This implies that the Clausius relation holds for the
Lovelock theory. In scalar-tensor gravity, we find, by using the
same procedure, the Clausius relation no longer holds. This
indicates that the apparent horizon of FRW universe in the
scalar-tensor gravity corresponds to a system of non-equilibrium
thermodynamics. We show this point by using the method developed
recently by Eling {\it et al.} for dealing with the $f(R)$ gravity.
\end{abstract}
\maketitle

\section{Introduction}
Quantum mechanics together with general relativity predicts that
black hole behaves like a black body, emitting thermal radiations,
with a temperature proportional to its surface gravity at the black
hole horizon and with an entropy proportional to its horizon area
\cite{Haw,Bek}. The Hawking temperature and horizon entropy together
with the black hole mass obey the first law of black hole
thermodynamics ($ dM=T dS$)~\cite{firstlaw}. The formulae of black
hole entropy and temperature have a certain universality in the
sense that the horizon area and surface gravity are purely geometric
quantities determined by the spacetime geometry.

Since the discovery of black hole thermodynamics in 1970's,
physicists have been speculating that there should be some relation
between the thermodynamic laws and Einstein equations. Otherwise,
how does general relativity know that the horizon area of black hole
is related to its entropy and the surface gravity to its temperature
\cite{Jac}? Indeed, Jacobson \cite{Jac} was able to derive Einstein
equations from the proportionality of entropy to the horizon area,
$A$, together with the fundamental relation (Clausius relation)
$\delta Q= T dS$, assuming the relation holds for all local Rindler
causal horizons through each spacetime point. Here $\delta Q$ and
$T$ are the variation of heat flow and Unruh temperature seen by an
accelerated observer just inside the horizon. More recently, Eling
{\it et al.} have found that one can not get the right equations of
motion for $f(R)$ gravity if one simply uses the Clausius relation
and the entropy assumption $S=\alpha f'(R)A$. In order to get the
equations of motion, an entropy production term has to be added to
the Clausius relation. They have argued that this corresponds to the
non-equilibrium thermodynamics of space-time~\cite{Jac1}. It is
interesting to see whether the non-equilibrium thermodynamics is
needed in other gravity theories. In this paper, we will discuss the
scalar-tensor gravity by following \cite{Jac1}. By using entropy
assumption $S=\alpha F(\phi)A$ in the scalar-tensor gravity and
Clausius relation with an appropriate entropy production term, we
can obtain correct equations of motion for the scalar-tensor
gravity. This suggests that for the scalar-tensor gravity the
non-equilibrium thermodynamics also has to be employed to derive the
dynamic equations of motion of space-times.

On the other hand, most discussions of black hole thermodynamics
have been focused on the stationary black holes.  For dynamical
(i.e., non-stationary) black holes,  Hayward has proposed a method
to deal with thermodynamics associated with trapping horizon of a
dynamic black hole in $4$-dimensional Einstein theory~\cite{Hayward,
Hayward1, Hayward2, Hayward3}. In this method, for spherical
symmetric space-times, Einstein equations can be rewritten in a form
called ``unified first law". Projecting this unified first law along
trapping horizon, one gets the first law of thermodynamics for
dynamical black hole. A definition of energy-supply, $\Psi$, is
introduced in the unified first law. It is an energy flux defined by
energy-momentum tensor of matter. After projecting along a vector
$\xi$ tangent to trapping horizon, one finds $\langle
A\Psi,\xi\rangle=\frac{\kappa}{8\pi G}\langle dA,\xi\rangle$. This
equation can be regarded as the Clausius relation of the dynamical
black hole. The Friedmann-Robertson-Walker (FRW) universe is one
kind of non-stationary spherically symmetric space-times. Certainly,
we can discuss its thermodynamics on the trapping horizon. However,
in the FRW universe, the trapping horizon (``outer trapping
horizon") is absent, instead there exists a kind of cosmological
horizon like trapping horizon (``inner trapping horizon"). This
horizon coincides with the apparent horizon in the context of the
FRW cosmology. In this paper, we therefore do not distinguish them.
We will focus on this apparent horizon and discuss associated
thermodynamics. An interesting question is whether the field
equations for non-Einstein gravity can be written to a form as the
unified first law in Einstein gravity. We do not try to resolve this
issue, instead we rewrite the field equations for non-Einstein
gravity to a form of the Einstein gravity by introducing an
effective energy-momentum tensor. Thus, we can use unified first law
to non-Einstein gravity theories and get the thermodynamics of
apparent horizon in the FRW universe in those theories. As a result,
we have the equation $\langle A\Psi,\xi\rangle=\frac{\kappa}{8\pi
G}\langle dA,\xi\rangle$ even for the non-Einstein gravity theories.
But, the energy-supply $\Psi$ here includes the contribution of the
effective energy-momentum tensor.

In order to get the heat $\delta Q$ (defined by integration of pure
matter energy flux) in the Clausius relation, we have to extract the
contribution of pure matter fields from the $\langle
A\Psi,\xi\rangle$ in the left hand side of the equation $\langle
A\Psi,\xi\rangle=\frac{\kappa}{8\pi G}\langle dA,\xi\rangle$ and put
the others into the right hand side of the equation. Now, does the
right hand side of the equation have the correct form of $T dS$ in
Clausius relation? If the answer is yes, the Clausius relation holds
and we are treating  equilibrium thermodynamics. On the contrary, if
the answer is no, the Clausius relation in equilibrium
thermodynamics does no longer hold, and we have to treat the system
with non-equilibrium thermodynamics. In this paper, we will use this
idea to treat the Lovelock gravity and scalar-tensor gravity. We
find that the Clausius relation holds in Lovelock gravity, while it
breaks for the scalar-tensor gravity.  We will also show this point
by using the method of Eling {\it et al.} developed for the $f(R)$
gravity~\cite{Jac1}.

There exist some works dealing with the relation between the
Einstein equations and the first law of thermodynamics. In a setup
of a special kind of spherically symmetric black hole spacetimes,
Padmanabhan {\it et al.}~\cite{Pad} showed that the Einstein
equations on the black hole horizon can be written into the first
law of thermodynamics, $dE=TdS -PdV$. This also holds in Lovelock
gravity~\cite{Pad}. In the setting of FRW universe, some authors
investigated the relation between the first law  and the Friedmann
equations describing the dynamic evolution of the
universe~\cite{Fri}. In particular, Cai and Kim in \cite{cai1}
derived the Friedmann equations by applying the fundamental relation
$\delta Q=T\delta S$ to the apparent horizon of the FRW universe
with any spatial curvature and assuming that the apparent horizon
has temperature and entropy
  \begin{equation}
  T = \frac{1}{2\pi R_A}, \ \ \  S = \frac{\pi R^2_A}{G},
  \end{equation}
  where $R_A$ is the apparent horizon radius. Further they showed
  that using the same procedure, the Friedmann equations can be
  also derived in the Gauss-Bonnet gravity and more general
  Lovelock gravity. For the scalar-tensor gravity and $f(R)$
  gravity, the possibility to derive the corresponding Friedmann
  equations in those theories was investigated in \cite{Akbar1}.
  More recently Akbar and Cai~\cite{Akbar2} have shown that at the apparent horizon, the
  Friedmann equation can be written into a form of the first law
  of thermodynamics with a volume change term, not only in
  Einstein gravity, but also in Lovelock gravity.

This paper is organized as follows. In Sec. II, we give a brief
review on the unified first law by generalizing it to
$(n+1)$-dimensional Einstein gravity. In Sec. III, we consider a FRW
universe and give the projection vector which will be used in the
following sections. In Sec. IV, we give the rigorous first law of
thermodynamics on the apparent horizon for the FRW universe in
Einstein theory. In Sec. V, we treat the thermodynamics of apparent
horizon in the Lovelock gravity and find that the Clausius relation
holds for the Lovelock gravity. In Sec. VI, we discuss the
scalar-tensor gravity and show that the Clausius relation no longer
holds.  In Sec. VII, we derive the equations of motion for the
scalar-tensor gravity by using the method of Eling {\it et al.}
developed for dealing with the $f(R)$ gravity. We end this paper
with conclusion in Sec.~VIII.

\section{A Brief Review on the Unified First Law}

Hayward has proposed a general definition of black hole dynamics on
trapping horizon in four dimensional Einstein theory~\cite{Hayward,
Hayward1, Hayward2, Hayward3}. In this section, we will make a brief
review and generalize his discussions to the $(n+1)$-dimensional
Einstein gravity.

For an arbitrary $(n+1)$-dimensional spherical symmetric space-time,
locally we can put its metric in the double-null form
\begin{equation}
ds^2 = -2e^{-f}d\xi^+ d\xi^-+r^2d\Omega_{n-1}^2\, ,
\end{equation}
where $d\Omega_{n-1}^2$ is the line element of an $(n-1)$-sphere
with unit radius, $r$ and $f$ are functions of $(\xi^+, \xi^-)$.
Certainly, there are some remainder freedoms to choose the
double-null coordinates. Assume that the space-time is
time-orientable, and $\partial_{\pm}=\partial /\partial \xi^{\pm}$
are future-pointing. Considering radial null geodesic congruence,
from $ds^2=0$, one can find that there are two kinds of null
geodesics corresponding to $\xi^+=\mathrm{constant}$ and
$\xi^{-}=\mathrm{constant}$, respectively. It is easy to get the
expansions of these two congruences
\begin{equation}
\theta_{\pm}=(n-1)\frac{\partial_{\pm}r}{r}\, .
\end{equation}
The expansion measures whether the light rays normal to the sphere
are diverging ($\theta_{\pm}>0$), or converging ($\theta_{\pm}<0$),
or equivalently, whether the area of the sphere is increasing or
decreasing in the null directions.

A sphere is said to be {\it trapped}, {\it untrapped} or {\it
marginal} if (on this sphere)
\begin{equation}\theta_{+}\theta_{-}>0\, ,\quad \theta_{+}\theta_{-}<0\, ,
\quad\mathrm{or} \quad \theta_{+}\theta_{-}=0\, .
\end{equation}
Note that
\begin{equation}
g^{ab}\partial_a r \partial_b
r=-\frac{2}{(n-1)^2}e^fr^2\theta_{+}\theta_{-}\, ,
\end{equation}
 we can write this definition to be
\begin{equation}
g^{ab}\partial_a r \partial_b r<0,\quad g^{ab}\partial_a r
\partial_b r>0,\quad g^{ab}\partial_a r \partial_b
r=0\, .
\end{equation}
If $e^f\theta_+\theta_-$ is a function with nonvanishing
derivatives, the space-time is divided into trapped and untrapped
regions, separated by marginal surface. Some subdivisions may be
made as follows.

\indent (i) A trapped sphere is said to be future, if
$\theta_{\pm}<0$ ($\partial_{\pm}r<0$) and past if $\theta_{\pm}
>0$ ($\partial_{\pm}r >0$).

\indent (ii) On an untrapped sphere, a spatial or null normal vector
$z$ is outgoing if $\langle dr,z\rangle>0$ and ingoing if $\langle
dr,z\rangle<0$. Equivalently, fixing the orientation by locally
$\theta_+>0$ and $\theta_-<0$, $z$ is outgoing if
$g(z,\partial_+)>0$ or $g(z,\partial_-)<0$ and ingoing if
$g(z,\partial_+)<0$ or $g(z,\partial_-)>0$. In particular,
$\partial_+$ and $\partial_-$ are, respectively, the outgoing and
ingoing null normal vectors.

\indent (iii) A marginal sphere with $\theta_+=0$ is future if
$\theta_{-}<0$, past if $\theta_->0$, bifurcating if $\theta_-=0$,
outer if $\partial_-\theta_+<0$, inner if $\partial_-\theta_+>0$ and
degenerate if $\partial_-\theta_+=0$.

\noindent The closure of a hypersurface foliated by future or past,
outer or inner marginal sphere is called a (nondegenerate) {\it
trapping horizon}.

In the works of Hayward, the future (past) outer trapping horizon is
taken as the definition of black (white) holes. i.e., on the
marginal sphere of trapping horizon, we have
\begin{equation}
\theta_{+}=0,\quad \theta_- <0,\quad \partial_-\theta_+<0\, ,
\end{equation}
or equivalently
\begin{equation}
\partial_{+}r=0,\quad \partial_- r<0,\quad \partial_-\partial_+r<0\,
.
\end{equation}
However, in the FRW universe, we will treat a horizon which is
similar to cosmological horizon in the de Sitter space-time. In this
case, the surface gravity is negative. So, we need not the
requirement of ``outer". In fact, we will take the future inner
trapping horizon as a system on which the thermodynamics will be
established.

The Misner-Sharp energy~\cite{MS,Hayward1,BR} is defined to be
\begin{equation}
\label{MisnerSharp} E=\frac{1}{16\pi G}(n-1)\Omega_{n-1}
r^{n-2}(1-g^{ab}\partial_a r\partial_b r)\, .
\end{equation}
This energy is the total energy (not only the passive energy) inside
the sphere with radius $r$. It is a pure geometric quantity. From
the definition, the ratio $E/r^{n-2}$ controls the formation of
black and white holes and trapped sphere
generally~\cite{Hayward1,BR}. There are a lot of definitions for
energy in general relativity, such as, ADM mass for asymptotically
flat space-time, Bondi-Sachs energy defined at null infinity of the
asymptotically flat space-time, Brown-York energy and Liu-Yau energy
{\it etc.}~\cite{ADM,Bondi,Sachs,BY,LY}. These definitions can be
found in a recent review~\cite{LBS}. The physical meanings of
Misner-Sharp energy, and the comparison of Misner-Sharp energy to
ADM mass and Bondi-Sachs energy have been given in
\cite{MS,Hayward1}. For spherical space-time, Brown-York energy
agrees with the Liu-Yau energy, but they both differ from the
Misner-Sharp energy. For example, for the four dimensional
Reissner-Nordstr$\mathrm{\ddot{o}}$m black hole, the Misner-Sharp
energy differs from the Brown-York or Liu-Yau mass by a term which
is the energy of the electromagnetic field inside the sphere. The
Misner-Sharp energy has the relation to the structure of the
space-time and one can relate it to Einstein equations (see
Eq.~(\ref{unfl}) below). This is an important advantage of
Misner-Sharp energy.

From the energy-momentum tensor $T_{ab}$, we can give two useful
invariants --- {\it work} and {\it energy-supply}:
\begin{equation}
W=-\frac{1}{2} \mathrm{trace} T = -g_{+-}T^{+-}\, ,
\end{equation}
\begin{equation}
\Psi_a=T_{a}{}^b\partial_b r+W\partial_a r\, .
\end{equation}

By using the Misner-Sharp energy and these two quantities, one can
find that the $(0,0)$ component of Einstein equations can be written
as
\begin{equation}
\label{unfl} dE=A\Psi+WdV \, ,
\end{equation}
where $A=\Omega_{n-1}r^{n-1}$ and $V=\frac{1}{n}\Omega_{n-1}r^{n}$
with $\Omega_{n-1}=2\pi^{n/2}/\Gamma(n/2)$ are area and volume of a
sphere with radius $r$. Equation~(\ref{unfl}) is called {\it unified
first law}. It is the natural result of Einstein equations for
spherical symmetric space-times.  The unified first law (\ref{unfl})
contains rich information.  After projecting it along different
directions, this equation gives different meanings.  For instance,
(i) projecting the unified first law along the future null infinity,
one has the Bondi energy loss equation; (ii) projecting the unified
first law along the flow of a thermodynamic material yields the
first law of relativistic thermodynamics; (iii) projecting the
unified first law along the trapping horizon, one obtaines the first
law of black hole thermodynamics. Here, we will concentrate on the
first law of black hole thermodynamics, which has the form
\begin{equation}
\label{flbh} \langle dE,z\rangle=\frac{\kappa}{8\pi G}\langle
dA,z\rangle+W \langle dV,z\rangle \, ,
\end{equation}
where $\kappa$ is the surface gravity of the trapping horizon, and
is defined by
\begin{equation}
\kappa=\frac{1}{2}\nabla^{a}\nabla_{a}r\, ,
\end{equation}
where $\nabla$ corresponds to the covariant derivative of two
dimension space normal to the sphere, and $z$ is a vector which is
tangent to the trapping horizon (Certainly, $z$ is not arbitrary and
must satisfy some conditions on the trapping horizon). In the double
null coordinates, $z$ can be expressed as
\begin{equation}
z=z^+ \partial_+ + z^{-}\partial_-\, .
\end{equation}
The equation~(\ref{flbh}) is obtained through projecting the unified
first law along the vector $z$. The nontrivial part is to show
\begin{equation}
\label{Proofentropy} \langle A\Psi,z\rangle=\frac{\kappa}{8\pi
G}\langle dA,z\rangle\, .
\end{equation}
As mentioned above, we take the horizon to be
\begin{equation}
\partial_{+}r=0\, .
\end{equation}
Then, one has on the marginal sphere,
\begin{equation}
z^a\partial_a (\partial_+
r)=z^+\partial_+\partial_+r+z^-\partial_-\partial_+ r=0\, .
\end{equation}
Using Einstein equations and the definition of surface gravity, one
can arrive at~(\ref{Proofentropy}). The most important thing is to
note that $z$ is not arbitrary, on the trapping horizon it must
satisfy the equation above, then
\begin{equation}
\label{eq19}
\frac{z^{-}}{z^{+}}=-\frac{\partial_+\partial_+r}{\partial_-\partial_+
r}\, .
\end{equation}
Therefore $z$ belongs to the one dimensional subspace of the tangent
space. The projection along $z$ replaces the extra differential by
the differential of state space for black hole,
 one can arrive at the exact first law of the trapping horizon
  thermodynamics of dynamic black holes~\cite{Hayward2}.
This equation (\ref{eq19}) plays a curial role to get the projection
vector. We will give this ratio for the FRW universe in the next
section.

\section{Trapping Horizon and Apparent Horizon of the FRW Universe}

Now, we consider an $(n+1)$-dimensional FRW universe. We  put the
FRW metric in the form
\begin{equation}
ds^2=h_{ab}dx^adx^b+\tilde{r}^2d\Omega_{n-1}^2
\end{equation}
where $x^0=t$, $x^1=r$, $\tilde{r}= a r$ is the radius of the sphere
and $a$ is the scale factor. It should be noted here, that
$\tilde{r}=\tilde{r}(t,r)$ plays the role of sphere radius $r$
defined in the previous section. Defining
\begin{equation}
d\xi^+=-\frac{1}{\sqrt{2}}\left(dt-\frac{a}{\sqrt{1-kr^2}}dr\right)\,
, \quad
d\xi^-=-\frac{1}{\sqrt{2}}\left(dt+\frac{a}{\sqrt{1-kr^2}}dr\right)\,
,
\end{equation}
where $k$ is the spatial curvature parameter of the FRW universe, we
can put the FRW metric into a double-null form
\begin{equation}
ds^2=-2 d\xi^+d\xi^- +\tilde{r}^2d\Omega_{n-1}^2\, .
\end{equation}
It is easy to find
\begin{equation}
\partial_+=\frac{\partial}{\partial\xi^+}=
-\sqrt{2}\left(\partial_t-\frac{\sqrt{1-kr^2}}{a}\partial_r\right)\,
,\quad
\partial_-=\frac{\partial}{\partial\xi^-}=
-\sqrt{2}\left(\partial_t+\frac{\sqrt{1-kr^2}}{a}\partial_r\right)\,
,
\end{equation}
where the minus signs ensure that $\partial_{\pm}$ are future
pointing. The trapping horizon, we denote it by $\tilde{r}_A$, is
defined to be
\begin{equation}
\partial_+ \tilde{r}|_{\tilde{r}=\tilde{r}_A}=0\, .
\end{equation}
Solving this equation, one finds
\begin{equation}
\label{r_A}
 \tilde{r}_A^2=\frac{1}{H^2+\frac{k}{a^2}}\, .
\end{equation}
This radius has the same form as apparent horizon~\cite{BR}. It is
not surprised because the trapping horizon and apparent horizon
coincide with each other in the FRW universe. On the other hand, we
have
\begin{equation}
\partial_- \tilde{r}|_{\tilde{r}=\tilde{r}_A}=- 2\tilde{r}_AH <0\, ,
\end{equation}
that is,  this trapping horizon is future. A similar calculation on
the trapping horizon gives
\begin{equation}
\label{zradio}
\partial_-
\partial_{+}\tilde{r}|_{\tilde{r}_A}=2\tilde{r}_A\left(\dot{H}+2H^2+\frac{k}{a^2}\right)\,
,\quad \partial_+
\partial_{+}\tilde{r}|_{\tilde{r}_A}=2\tilde{r}_A\left(\dot{H}-\frac{k}{a^2}\right)\,
.
\end{equation}
By definition, one can find the surface gravity
\begin{equation}
\kappa=-\frac{\tilde{r}_A}{2}\left(\dot{H}+2H^2+\frac{k}{a^2}\right)=
-\frac{1}{\tilde{r}_A}\left(1-\frac{\dot{\tilde{r}}_A}{2H\tilde{r}_A}\right)\,
.
\end{equation}
Further, we define
\begin{equation}
\epsilon=\frac{\dot{\tilde{r}}_A}{2H\tilde{r}_A}\, .
\end{equation}
Here, we assume $\epsilon<1$ such that $\kappa<0$. In another words,
we are treating an ``inner" trapping horizon, rather than ``outer "
trapping horizon (with positive surface gravity) discussed by
Hayward. Note that in references~\cite{cai1,Fri}, in fact, an
approximation $\epsilon \ll 1$ has been used in calculating the
energy flow crossing the apparent horizon. In the present paper, no
any approximation will be used. In terms of the horizon radius
$\tilde r_A$,  we have
\begin{equation}
\dot{H}-\frac{k}{a^2}=-\frac{2\epsilon}{\tilde{r}_A^2}\, .
\end{equation}
Substituting this into~(\ref{zradio}), we get
\begin{equation}
\frac{z^{-}}{z^{+}}=-\frac{\partial_+\partial_+\tilde{r}}{\partial_-\partial_+
\tilde{r}}|_{\tilde{r}_A}=\frac{\epsilon}{1-\epsilon}\, .
\end{equation}
Let $z^{+}=1$, then $z^{-}=\frac{\epsilon}{1-\epsilon}$, in the
coordinates $(t,r)$, we then can express $z$ as
\begin{equation}
z=-\frac{\sqrt{2}}{1-\epsilon}\left(\frac{\partial}{\partial
t}-(1-2\epsilon)Hr\frac{\partial}{\partial r}\right)\, ,
\end{equation}
where we have used the relation
\begin{equation}
\frac{\sqrt{1-kr^2}}{a}|_{\tilde{r}_A}=Hr|_{\tilde{r}_A}\, ,
\end{equation}
 on the trapping horizon. Thus, if we use the $(t, r)$
coordinates instead of double-null coordinates, any project vector
$\xi$ must have the form
\begin{equation}
\label{projectvector} \xi=\xi^t\left(\frac{\partial}{\partial
t}-(1-2\epsilon)Hr\frac{\partial}{\partial r}\right)\, .
\end{equation}
Certainly, we can choose $\xi^t=1$. All calculations in this section
are pure geometrical.  Therefore the results in this section are
applicable for the FRW universe in any gravity theory.


\section{Thermodynamics of Apparent Horizon in Einstein Gravity}

Consider a Lagrangian for an $(n+1)$-dimensional Einstein gravity
with perfect fluid
\begin{equation}
\mathcal{L}=\frac{1}{16\pi G}R+\stackrel{m}{\mathcal{L}}\, ,
\end{equation}
where $\stackrel{m}{\mathcal{L}}$ denotes the Lagrangian for the
perfect fluid.  In the FRW universe, the energy-momentum tensor of
the perfect fluid  has the form
\begin{equation}
T_{ab}=\stackrel{m}{T_{ab}}= (\rho_m+p_m)U_aU_b+ p_mg_{ab}\, ,
\end{equation}
where  $\rho_m$ and $p_m$ are the energy density and pressure of the
perfect fluid, respectively. It is easy to find that the
energy-momentum tensor projecting onto two dimensional space-time
normal to the sphere has the form
\begin{equation}
T_{ab}={\rm diag}(\rho_{m},  \frac{p_{m} a^2}{1-kr^2}),\quad
T_{a}{}^{b}={\rm diag} (-\rho_m, p_m).
\end{equation}
 and then the work term and energy-supply are
\begin{eqnarray}
 && W=\frac{1}{2}(\rho_m-p_{m})\, , \\
&& \Psi_t=-\frac{1}{2}(\rho_{m}+p_m)H\tilde{r}\, , \\
&& \Psi_r=\frac{1}{2}(\rho_{m}+p_m)a\, ,
\end{eqnarray}
respectively. One then has
\begin{equation}
\Psi=\Psi_t dt+\Psi_r
dr=-\frac{1}{2}(\rho_{m}+p_m)H\tilde{r}dt+\frac{1}{2}(\rho_{m}+p_m)a
dr\, .
\end{equation}
Thus, on the trapping horizon/apparent horizon we have
\begin{eqnarray}
\label{dE}
d E&=&A \Psi +W dV=A \Psi +A W d \tilde {r}_A\nonumber\\
 &=&-A(\rho_{m}+p_m )H\tilde{r}_Adt+A \rho_{m} d\tilde{r}_A\nonumber \\
 &=&V d\rho_{m}+\rho_m dV=d\left(\rho_m V\right)\, .
\end{eqnarray}
Substituting the first Friedmann equation ($H^2+k/a^2= 16\pi G
\rho_m/(n(n-1)))$ into the last line in the above equation, one can
get nothing but the differential of the Misner-Sharp energy. On the
other hand, if use the Misner-Sharp energy (\ref{MisnerSharp})
inside the apparent horizon with the radius (\ref{r_A}), the above
equation (\ref{dE}) tells us nothing, but the first Friedmann
equation~\cite{Akbar2}.

Let $\xi$ be a vector tangent to the apparent horizon,  which can be
expressed as~(\ref{projectvector}).  From now on we choose
$\xi^{t}=1$, thus we have
\begin{equation}
\langle d E, \xi\rangle=- A H \tilde{r}_A\left[ (1-2\epsilon)\rho_m
+p_m)\right].
\end{equation}
On the other hand,
\begin{equation}
\frac{\kappa}{8\pi G}dA=-(n-1)\left(1-\epsilon\right)\frac{A}{8\pi
G\tilde{r}_A^2}(H\tilde{r}_Adt+adr)\, .
\end{equation}
By using Friedmann equation, we have
\begin{eqnarray}
\frac{\kappa}{8\pi G}\langle dA,\xi \rangle &=&-(n-1)
2\epsilon\left(1-\epsilon\right)\frac{A}{8\pi
G\tilde{r}_A^2} H\tilde{r}_A\nonumber \\
&=&- (1-\epsilon) AH\tilde{r}_A(\rho_m+p_m)\, .
\end{eqnarray}
Similarly, one can show
\begin{equation}
\langle WdV, \xi\rangle=\epsilon AH\tilde{r}_A(\rho_m-p_m)\, .
\end{equation}
Combining them yields
\begin{equation}
\frac{\kappa}{8\pi G}\langle dA,\xi \rangle + \langle WdV,
\xi\rangle=- A H \tilde{r}_A \left[ (1-2\epsilon)\rho_m
+p_m)\right]\, .
\end{equation}
Thus we have shown that the unified first law on the inner trapping
horizon has the form
\begin{equation}
\langle d E, \xi\rangle=\frac{\kappa}{8\pi G}\langle dA,\xi \rangle
+ \langle WdV, \xi\rangle.
\end{equation}
Here some remarks are in order. (i) The unified first law
(\ref{unfl}) is not a real first law of thermodynamics, but just an
identity concerning the $(0,0)$ component of Einstein equations.
However, the projection of the unified first law along a trapping
horizon (or apparent horizon in FRW cosmology context) gives a real
first law of thermodynamics. (ii) The Misner-Sharp energy plays an
important role in the unified first law and the definition of work
and energy-supply is very useful. The separation of work and
energy-supply gives a very similar form as the first law of
thermodynamics before projecting it along the horizon. (iii) On the
horizon, the energy-supply has the form
\begin{equation}
\langle A\Psi, \xi\rangle=\frac{\kappa}{8\pi G}\langle dA,\xi
\rangle
\end{equation}
This is the Clausius relation in the version of black hole
thermodynamics. The left hand side of the above equation is nothing
but heat flow $\delta Q$ defined by the matter energy-momentum
tensor. The right hand side has the form  $T d S$ if one identities
that temperature $T=\kappa/2\pi$ and $S=A/4G$. Thus we conclude that
in Einstein theory, the unified first law also imples the Clausius
relation $\delta Q=T  dS$. (iv) For the FRW cosmology in Einstein
theory, the Misner-Sharp energy inside the trapping horizon is
$E=\rho_m V$ (see (\ref{dE}))---the energy density times the volume.
This energy form for the FRW cosmology in Einstein theory is very
special. In general, the Misner-Sharp energy has no such a form in
other gravity theories, which will be seen in next sections.


\section{Thermodynamics of Apparent Horizon in Lovelock Gravity}

The Lagrangian of the Lovelock gravity~\cite{Lovelock} consists of
the dimensionally extended Euler densities
\begin{equation}
\mathcal{L}=\sum_{i=0}^{m}c_i\mathcal{L}_i\, ,
\end{equation}
where $c_i$ is an arbitrary constant, $m\le [n/2]$ and
$\mathcal{L}_i$ is the Euler density of a $2i$-dimensional manifold
\begin{equation}
\mathcal{L}_i=\frac{1}{2^i} \delta_{c_1 d_1 \cdots c_i d_i}^{a_1
b_1\cdots a_i b_i} R^{c_1 d_1}{}{}_{a_1 b_1} \cdots R^{c_i
d_i}{}{}_{a_i b_i}\, .
\end{equation}
$\mathcal{L}_0$ corresponds to the cosmological term.
$\mathcal{L}_1$ is just the Einstein-Hilbert term, and
$\mathcal{L}_2$ corresponds to the so called ``Gauss-Bonnet" term.
Although the Lagrangian of the Lovelock theory contains higher
curvature terms, there are no terms with more than second order
derivatives of metric in equations of motion. This point can be
directly found from the equations of motion
\begin{equation}
\mathcal{G}^{a}_b=\sum_{i=0}^{m}\frac{c_i}{2^{i+1}} \delta_{b c_1
d_1 \cdots c_i d_i}^{a a_1 b_1\cdots a_i b_i} R^{c_1 d_1}{}{}_{a_1
b_1} \cdots R^{c_i d_i}{}{}_{a_i b_i}=0\, .
\end{equation}
If we introduce matter fields, the equations of motion become
\begin{equation}
\mathcal{G}^{a}_b=\sum_{i=0}^{m}\frac{c_i}{2^{i+1}} \delta_{b c_1
d_1 \cdots c_i d_i}^{a a_1 b_1\cdots a_i b_i} R^{c_1 d_1}{}{}_{a_1
b_1} \cdots R^{c_i d_i}{}{}_{a_i b_i}=8\pi G {\stackrel{m}{T_b^a}}\,
.
\end{equation}
In the FRW cosmology, the energy-momentum tensor is still taken to
be that of perfect fluid.   We can put this equations of motion into
the standard form in Einstein gravity by moving those terms except
the Einstein tensor into the right hand side of the equation
\begin{equation}
G^a_b=8\pi G \left({\stackrel{m}{T_b^a}}
+{\stackrel{e}{T_b^a}}\right)\, ,
\end{equation}
where the effective energy-momentum tensor ${\stackrel{e}{T_b^a}}$
has the expression
\begin{equation}
\label{T_eff}
 {\stackrel{e}{T_b^a}}=-\frac{1}{8\pi
G}\sum_{i=0,i\neq 1}^{m}\frac{c_i}{2^{i+1}} \delta_{b c_1 d_1 \cdots
c_i d_i}^{a a_1 b_1\cdots a_i b_i} R^{c_1 d_1}{}{}_{a_1 b_1} \cdots
R^{c_i d_i}{}{}_{a_i b_i}\, .
\end{equation}
In the FRW metric, some non-vanishing components of Riemann tensor
have the forms
\begin{eqnarray}
 &&R^{tr}{}{}_{tr}=\dot{H}+H^2,\quad
R^{ti}{}{}_{tj}=\left(\dot{H}+H^2\right)\delta^i_j,\nonumber \\
&& R^{ri}{}{}_{rj}=\left(H^2+\frac{k}{a^2}\right)\delta^i_j,\quad
R^{ij}{}{}_{kl}=\left(H^2+\frac{k}{a^2}\right)\delta^{ij}_{kl}.
\end{eqnarray}
Substituting these into (\ref{T_eff}), we have
\begin{equation}
{\stackrel{e}{T_t^t}}=-\frac{1}{8\pi G}\sum_{i=0,i\neq
1}^{m}\frac{\widehat{c}_i}{2}\left(H^2+\frac{k}{a^2}\right)^i\, ,
\end{equation}
\begin{equation}
{\stackrel{e}{T_r^r}}=-\frac{1}{8\pi G}\sum_{i=0,i\neq
1}^{m}\frac{\widehat{c}_i}{2}\left[\left(H^2+\frac{k}{a^2}\right)^i
+\frac{2i}{n}\left(H^2+\frac{k}{a^2}\right)^{i-1}\left(\dot{H}-\frac{k}{a^2}\right)\right]\,
,
\end{equation}
where
\begin{equation}
\widehat{c}_i=\frac{n!}{(n-2i)!}c_i\, .
\end{equation}
The work term can be decomposed as
\begin{equation}
W=\stackrel{m}{W}+\stackrel{e}{W}\, ,
\end{equation}
where
\begin{equation}
\stackrel{m}{W}=-\frac{1}{2}h^{ab}\stackrel{m}{T}_{ab}=\frac{1}{2}(\rho_m-p_m)\,
,\quad \stackrel{e}{W}=-\frac{1}{2}h^{ab}\stackrel{e}{T}_{ab}\, .
\end{equation}
The higher curvature terms produce the effective work term
\begin{equation}
{\stackrel{e}{W}}=\frac{1}{8\pi G}\sum_{i=0,i\neq
1}^{m}\frac{\widehat{c}_i}{2}\left[\left(H^2+\frac{k}{a^2}\right)^i
+\frac{i}{n}\left(H^2+\frac{k}{a^2}\right)^{i-1}\left(\dot{H}-\frac{k}{a^2}\right)\right]\,
.
\end{equation}
Similarly, the energy-supply $\Psi$ can also be divided into
\begin{equation}
\Psi={\stackrel{m}{\Psi}}+{\stackrel{e}{\Psi}}\, ,
\end{equation}
where
\begin{equation}
{\stackrel{m}{\Psi}}_a={\stackrel{m}{T_a^b}}\partial_b
\tilde{r}+{\stackrel{m}{W}}\partial _a \tilde{r}\, , \quad
{\stackrel{e}{\Psi}}_a={\stackrel{e}{T_a^b}}\partial_b
\tilde{r}+{\stackrel{e}{W}}\partial _a \tilde{r}\, .
\end{equation}
After some calculations we arrive at
\begin{equation}
{\stackrel{m}{\Psi}}=-\frac{1}{2}(\rho_m+p_m)H\tilde{r}dt+\frac{1}{2}(\rho_m+p_m)a
dr\, ,
\end{equation}
\begin{eqnarray}
{\stackrel{e}{\Psi}}&=&\frac{1}{16\pi G}\sum_{i=0,i\neq
1}^{m}\left[\widehat{c}_i\frac{i}{n}\left(H^2+\frac{k}
{a^2}\right)^{i-1}\left(\dot{H}-\frac{k}{a^2}\right)\right]H\tilde{r}dt\nonumber
\\
&&-\frac{1}{16\pi G}\sum_{i=0,i\neq
1}^{m}\left[\widehat{c}_i\frac{i}{n}\left(H^2+\frac{k}
{a^2}\right)^{i-1}\left(\dot{H}-\frac{k}{a^2}\right)\right]a dr\, .
\end{eqnarray}
Thus, we have put the Lovelock theory into the form of Einstein
theory with an effective energy-momentum tensor, and the effective
energy-momentum tensor has been used in the work and energy-supply
terms. These imply that the unified first law discussed previously
is applicable as well here and the energy has the form of the
Misner-Sharp energy~(\ref{MisnerSharp}).

 The Friedmann equations of Lovelock gravity read~\cite{cai1}
\begin{equation}
\sum_{i=0}^{m}\widehat{c}_i\left(H^2+\frac{k}{a^2}\right)^i= 16 \pi
G \rho_m\, ,
\end{equation}
\begin{equation}
\sum_{i=0 }^{m}i
\widehat{c}_i\left(H^2+\frac{k}{a^2}\right)^{i-1}\left(\dot{H}-\frac{k}{a^2}\right)=-n8\pi
G (\rho_m+p_m)\, .
\end{equation}
Using the Friedmann equations and the work and energy-supply terms
given above, one can obtain the differential of the Misner-Sharp
energy from the unified first law. Turn around, if we use the
Misner-Sharp energy and work and energy-supply terms, we can obtain
the Friedmann equations in the Lovelock gravity from the unified
first law.  An interesting point is here that for the Lovelock
gravity, we can not put the Misner-Sharp energy inside the trapping
horizon into the form $E=\rho_m V$ since the effective
energy-momentum tensor will make a contribution to the energy inside
the horizon~\cite{Akbar2}. Projecting the unified first law along
the trapping horizon, we have
\begin{equation}
\langle d E, \xi\rangle=\frac{\kappa}{8\pi G}\langle dA,\xi \rangle
+ \langle WdV, \xi\rangle\, ,
\end{equation}
since we have written the equation of motion for the Lovelock
gravity to the form of Einstein gravity. This projection implies
that we have
\begin{equation}
\langle A\Psi, \xi \rangle=\frac{\kappa}{8\pi G}\langle dA,\xi
\rangle\, .
\end{equation}
Namely, we have
\begin{equation}
\langle A{\stackrel{m}{\Psi}}, \xi \rangle=\frac{\kappa}{8\pi
G}\langle dA,\xi \rangle-\langle A{\stackrel{e}{\Psi}}, \xi
\rangle\, .
\end{equation}
Clearly, the left hand side is just the ``energy-supply" projecting
along $\xi$, which is nothing, but the heat flow $\delta Q$ defined
by pure matter energy-momentum tensor. Thus, an interesting question
is whether the right hand side of the equation can be of the form
$TdS$ in the Clausius relation as the case in Einstein gravity? The
answer is yes: the right hand side of the equation can indeed be
written to a form with the surface gravity times the differential of
the entropy in Lovelock theory projecting along $\xi$.  Now we show
this. From the definition of $\stackrel{e}{\Psi}$, we have
\begin{eqnarray}
\frac{\kappa}{8\pi G}\langle d A,\xi \rangle-\langle
A{\stackrel{e}{\Psi}}, \xi \rangle&=&- \frac{A}{4\pi
G\tilde{r}_A^2}\epsilon(1-\epsilon)(n-1)H\tilde{r}_A\nonumber\\
&-&\frac{A}{4\pi G}\epsilon(1-\epsilon)\sum_{i=0,i\neq 1}^{m} \left[
c_i\frac{i(n-1)!}{(n-2i)!}\left(\frac{1}{\tilde{r}_A^2}\right)^{i}\right]H\tilde{r}_A\,
.
\end{eqnarray}
Having considered $c_1=1$,  we can rewrite the above equation into
\begin{eqnarray}
\frac{\kappa}{8\pi G}\langle d A,\xi \rangle-\langle
A{\stackrel{e}{\Psi}}, \xi \rangle&=& -\frac{A}{4\pi
G}\epsilon(1-\epsilon)\sum_{i=0}^{m} \left[
c_i\frac{i(n-1)!}{(n-2i)!}\left(\frac{1}{\tilde{r}_A^2}\right)^{i}\right]H\tilde{r}_A\nonumber
\\
&=&-\frac{1}{2\pi\tilde{r}_A}(1-\epsilon)\Omega_{n-1}\langle
\frac{1}{4G}\sum_{i=0}^{m} \left[
c_i\frac{i(n-1)!}{(n-2i)!}\tilde{r}_A^{n-2i}\right]d\tilde{r}_A
,\xi \rangle\nonumber \\
&=&\frac{\kappa}{2\pi}\Omega_{n-1}\langle \frac{1}{4G}\sum_{i=0}^{m}
\left[ c_i\frac{i(n-1)!}{(n-2i+1)!}~d\tilde{r}_A^{n-2i+1}\right]
,\xi
\rangle\nonumber \\
&=&\frac{\kappa}{2\pi}\langle d\left\{\frac{A}{4G}\sum_{i=0}^{m}
\left[
c_i\frac{i(n-1)!}{(n-2i+1)!}~\tilde{r}_A^{2-2i}\right]\right\}
,\xi \rangle\nonumber \\
&=&T\langle dS, \xi\rangle \, ,
\end{eqnarray}
where $T=\kappa /2\pi$ and
\begin{equation}
S=\frac{A}{4G}\sum_{i=0}^{m} \left[
c_i\frac{i(n-1)!}{(n-2i+1)!}~\tilde{r}_A^{2-2i}\right]\, .
\end{equation}
Thus we have shown that $\frac{\kappa}{8\pi G}\langle d A,\xi
\rangle-\langle A{\stackrel{e}{\Psi}}, \xi \rangle$ is exactly the
surface gravity times a total differential projecting along the
tangent direction of the trapping horizon. This total differential
is nothing but the differential of the horizon entropy defined in
Lovelock gravity~\cite{cai2,cai1,Akbar2}.

Above discussions tell us: In Lovelock gravity, if one uses the pure
matter energy-momentum to define $\delta Q$, i.e., $\delta Q=
\langle A{\stackrel{m}{\Psi}},\xi \rangle$, then $\frac{\kappa}{8\pi
G}\langle d A,\xi \rangle-\langle A{\stackrel{e}{\Psi}}, \xi
\rangle$ is of the form $TdS$. That is, {\it the Clausius relation
$\delta Q= TdS$ still holds in Lovelock gravity}.

An interesting question arises: does the Clausius relation always
hold for any gravity theory?  In the next section, using the method
developed in this section, we will show that the Clausius relation
does not hold in the scalar-tensor theory.

\section{Thermodynamics of Apparent Horizon in Scalar-Tensor Gravity}

In the Jordan frame, the Lagrangian of the scalar-tensor gravity in
$(n+1)$-dimensional space-times can be written as
\begin{equation}
\label{scalar-tensor}
 \mathcal{L}=\frac{1}{16\pi G}
F(\phi)R-\frac{1}{2}g^{ab}\partial_{a}\phi\partial_{b}\phi-V(\phi)+\mathcal{L}_{m}\,
,
\end{equation}
where $F(\phi)$ is an positive continuous function of the scalar
field $\phi$ and $V(\phi)$ is its potential. Varying the action, we
have the equations of motion
\begin{equation}
F G_{ab}+g_{ab}\nabla^2F-\nabla_{a}\nabla_{b}F=8 \pi G
\Big{(}\stackrel{\phi}{T_{ab}}+\stackrel{m}{T_{ab}}\Big{)}\, ,
\end{equation}
\begin{equation}
\nabla^2\phi-V^{'}(\phi)+\frac{1}{16\pi G}F^{'}(\phi)R=0\, ,
\end{equation}
where $\stackrel{m}{T_{ab}}$ is the energy-momentum tensor of
matter. We denote $\stackrel{\phi}{T_{ab}}$ by
\begin{equation}
\stackrel{\phi}{T_{ab}}=\partial_a\phi\partial_b\phi
-g_{ab}\left(\frac{1}{2}g^{cd}\partial_c\phi\partial_d\phi+V(\phi)\right)\,
.
\end{equation}
Note that here $\stackrel{\phi}{T_{ab}}$ is not the energy-momentum
tensor of the scalar field.  As the case of Lovelock gravity, in
order to use the unified first law in Einstein gravity, we rewrite
the equations of motion into the following form
\begin{equation}
G_{ab}=8 \pi G T_{ab}=8 \pi
G\frac{1}{F}\Big{(}\stackrel{\phi}{T_{ab}}+\stackrel{m}{T_{ab}}+\stackrel{e}{T_{ab}}\Big{)}\,
,
\end{equation}
where
\begin{equation}
\stackrel{e}{T_{ab}} =\frac{1}{8\pi
G}\left(-g_{ab}\nabla^2F+\nabla_{a}\nabla_{b}F\right)\, .
\end{equation}
In the FRW metric, it is easy to find that $\nabla^2 F$ and the
non-vanishing components of $\nabla_a\nabla_b F$ are
\begin{equation}
\nabla^2 F=-\ddot{F}-nH\dot{F}\,, \quad \nabla_t \nabla_t
F=\ddot{F},\quad\nabla_r\nabla_rF=-\frac{a^2}{1-kr^2}H\dot{F}\, ,
\end{equation}
respectively.  We then have
\begin{equation}
\stackrel{e}{T_{t}^{t}}=\frac{1}{8\pi G}n H \dot{F}\, , \quad
\stackrel{e}{T_{r}^{r}}=\frac{1}{8\pi
G}\left(\ddot{F}+(n-1)H\dot{F}\right)\, .
\end{equation}
The work term can be decomposed as
\begin{equation}
\label{W}
 W=\stackrel{\phi}{W}+\stackrel{m}{W}+\stackrel{e}{W}\, ,
\end{equation}
with
\begin{equation}
\stackrel{\phi}{W}+\stackrel{m}{W}=\frac{1}{2F}(\rho_{\phi}+\rho_m-p_{\phi}-p_m)\,
,
\end{equation}
\begin{equation}
\stackrel{e}{W}=-\frac{1}{16\pi
GF}\left(\ddot{F}+(2n-1)H\dot{F}\right)\, ,
\end{equation}
where
\begin{equation}
\label{densitypresure}
 \rho_\phi=\frac{1}{2}\dot \phi^2 +V(\phi),
\ \ \ p_\phi= \frac{1}{2}\dot \phi^2 -V(\phi).
\end{equation}
  Similarly, the
energy-supply have the form
\begin{equation}
\label{P}
\Psi=\stackrel{\phi}{\Psi}+\stackrel{m}{\Psi}+\stackrel{e}{\Psi}\, ,
\end{equation}
with
\begin{equation}
\stackrel{\phi}{\Psi}+\stackrel{m}{\Psi}=-\frac{1}{2F}(\rho_{\phi}+\rho_m+
p_{\phi}+p_m)H\tilde{r}dt
+\frac{1}{2F}(\rho_{\phi}+\rho_m+p_{\phi}+p_m)adr\, ,
\end{equation}
\begin{equation}
\stackrel{e}{\Psi}=-\frac{1}{16\pi G
F}\left(\ddot{F}-H\dot{F}\right)H\tilde{r}dt+\frac{1}{16\pi
GF}\left(\ddot{F}-H\dot{F}\right)adr\, .
\end{equation}
On the trapping horizon/apparent horizon, the unified first law
tells us
\begin{eqnarray}
\label{STufirstlaw}
d E&=&A \Psi +W dV=A \Psi +A W d\tilde {r}_A\nonumber\\
&=&\frac{A}{F}\Bigg{[}-(\rho_{\phi}+\rho_m+p_{\phi}+p_m
)H\tilde{r}_Adt+
 (\rho_{\phi}+\rho_m) d\tilde{r}_A\nonumber \\
&-&\frac{1}{8\pi G}\left(\ddot{F}-H\dot{F}\right)H\tilde{r}_Adt
-\frac{1}{8\pi G}n H\dot{F}d\tilde{r}_A\Bigg{]}\, .
\end{eqnarray}
By using Friedmann equations, one can find that $dE$ is nothing but
the exterior differential of the Misner-Sharp energy
\begin{equation}
E=\frac{1}{16\pi
G}(n-1)\Omega_{n-1}\tilde{r}_A^{n-2}=\frac{V}{F}\left(\rho_{\phi}+\rho_m-\frac{1}{8\pi
G}nH\dot{F}\right)\, .
\end{equation}
On the other hand, if substituting the Misner-Sharp energy, work and
energy-supply terms defined above into the unified first
law~(\ref{STufirstlaw}), we can obtain the Friedmann equations in
the Scalar-Tensor gravity~\cite{Akbar1}:
\begin{equation}
\label{FW1} \frac{1}{2} n (n-1)F\left(H^2+\frac{k}{a^2}\right)+nH
\dot{F}=8\pi G \left(\rho_{\phi}+\rho_{m}\right)\, ,
\end{equation}
\begin{equation}
\label{FW2}
-(n-1)F\left(\dot{H}-\frac{k}{a^2}\right)-(\ddot{F}-H\dot{F}) =8\pi
G \left(\rho_{\phi}+p_{\phi}+\rho_m+p_m \right)\, ,
\end{equation}
where $\rho_{\phi}$ and $p_{\phi}$ have the same form
as~(\ref{densitypresure}). Since we have rewritten the equations of
motion of the scalar-tensor gravity into a form as in Einstein
theory, the following equation should hold
\begin{equation}
\langle d E, \xi\rangle=\frac{\kappa}{8\pi G}\langle d A,
\xi\rangle+\langle WdV, \xi\rangle\, .
\end{equation}
where $W$ is given by (\ref{W}).  Further we have
\begin{equation}
\langle A\Psi, \xi \rangle=\frac{\kappa}{8\pi G}\langle dA,\xi
\rangle\, ,
\end{equation}
where $\Psi$ is given by (\ref{P}).  Substituting $\Psi$ into the
above equation,  we have
\begin{equation}
\label{APsi} F \langle A\stackrel{\phi}{\Psi}, \xi \rangle+F\langle
A\stackrel{m}{\Psi}, \xi \rangle =\frac{\kappa F}{8\pi G}\langle
dA,\xi \rangle-F \langle A\stackrel{e}{\Psi}, \xi \rangle\, .
\end{equation}
The left hand side of this equation can be explicitly expressed as
\begin{equation}
F \langle A\stackrel{\phi}{\Psi}, \xi \rangle+F\langle
A\stackrel{m}{\Psi}, \xi
\rangle=\langle-\frac{1}{2}(\rho_{\phi}+\rho_m+p_{\phi}+p_m)H\tilde{r}dt
+\frac{1}{2}(\rho_{\phi}+\rho_m+p_{\phi}+p_m)adr,\xi\rangle\, .
\end{equation}
This is just the energy-supply term provided by matter and the
scalar field. We denote it by $\delta Q$ as the case in Einstein
gravity.  Can we put the right hand side of equation (\ref{APsi})
into a form with the surface gravity times a total differential
projecting along the vector $\xi$ as the case of Lovelock gravity?
The answer is no in the present case.  To see this, let us note that
the right hand side of the equation can be expressed as
\begin{eqnarray}
\frac{\kappa F}{8\pi G}\langle dA,\xi \rangle-F\langle
A\stackrel{e}{\Psi}, \xi \rangle &=&\frac{\kappa F}{8\pi G
\tilde{r}_A}A(n-1)2\epsilon H\tilde{r}_A-\frac{\kappa }{8\pi G
}A \tilde{r}_A\left(\ddot{F}-H\dot{F}\right)H\tilde{r}_A\nonumber \\
&=&T\langle dS,\xi \rangle+T \frac{A}{4 G}
\tilde{r}_A^2\left(\frac{\dot{F}}{\tilde{r}_A^2}-H\ddot{F}+H^2\dot{F}\right)\,
,
\end{eqnarray}
where
\begin{equation}
T=\frac{\kappa}{2\pi}\, , \quad S=\frac{F(\phi) A}{4G}\, .
\end{equation}
Here $S$ has the form of the entropy of black holes in the
scalar-tensor gravity~\cite{cai3}. Thus equation (\ref{APsi}) can be
reexpressed as
\begin{equation}
\label{eqdQ}
  \delta Q =T dS+Td _iS\, ,
\end{equation}
where
\begin{equation}
\label{eqd_iS} d_iS=\frac{A}{4 G}
\tilde{r}_A^2\left(\frac{\dot{F}}{\tilde{r}_A^2}-H\ddot{F}+H^2\dot{F}\right)\,
.
\end{equation}
The equation (\ref{eqdQ}) implies that the Clausius relation $\delta
Q= T dS$ does not hold for the scalar-tensor gravity. The term $d _i
S$ in (\ref{eqd_iS}) can be interpreted as the entropy production
term in the non-equilibrium thermodynamics associated with the
apparent horizon.  Indeed, in Einstein gravity, Jacobson~\cite{Jac}
used the Clausius relation $\delta Q= Td S$ and derived the Einstein
field equations. However, recently Eling {\it et al.}~\cite{Jac1}
have found that the Clausius relation does not hold for the $f(R)$
gravity, and that in order to obtain the equations of motion for the
$f(R)$ gravity, an entropy production term has to be added to the
Clausius relation like (\ref{eqdQ}). In the next section, following
\cite{Jac,Jac1}, we will show that indeed for the scalar-tensor
gravity, an additional entropy production term is needed for
deriving the equations of motion.

\section{Scalar-Tensor Gravity and Non-equilibrium Thermodynamics}

In \cite{Jac}, Jacobson  derived Einstein equations from the
proportionality of entropy to the horizon area, $A$, together with
the fundamental Clausius relation $\delta Q= TdS$, assuming that the
relation holds for all local Rindler causal horizons through each
space-time point. Here $\delta Q$ and $T$ are the variation of heat
and Unruh temperature seen by a accelerated observer just inside the
horizon. Recently, Eling{\it et al.}~\cite{Jac1} have shown that the
Clausius relation plus the entropy assumption $S=\alpha A f'(R)$ can
not give the correct equations of motion for the $f(R)$ gravity. In
order to get correct equations of motion, one has to modify the
equilibrium Clausius relation to a non-equilibrium one; an entropy
production term needs to be added to the Clausius relation of
equilibrium thermodynamics. Namely, the $f(R)$ gravity corresponds
to a non-equilibrium thermodynamics of space-time. In this section,
we will deal with the scalar-tensor gravity by using their method.

For a space-time point $p$ in $(n+1)$-dimensions, locally, one can
define a causal horizon as in \cite{Jac1}: Choose a spacelike
$(n-1)$-surface patch $B$ including $p$ and the choose one side of
the boundary of the past of $B$. Near the point $p$, this boundary
is a congruence of the null geodesics orthogonal to $B$. These
comprise the horizon.  To define the heat flux, we can employ an
approximate boost killing vector $\chi$ which is future pointing on
the causal horizon and vanishes at $p$. $\chi$ has a relation with
the tangent vector of causal horizon $k$: $\chi=-\lambda k$, where
$\lambda$ is the affine parameter of the corresponding null geodesic
line. The heat is defined to be the boost energy current of matter
(including the scalar field in the scalar-tensor gravity) across the
horizon
\begin{equation}
\delta Q=\int T_{ab}\chi^a d\Sigma^b\, .
\end{equation}
Using relation $\chi=-\lambda k$, and temperature $T=\hbar/2\pi$, we
have
\begin{equation}
\frac{\delta Q}{T}=\frac{2\pi}{\hbar}\int
T_{ab}k^ak^b(-\lambda)d\lambda d^{n-1}A\, ,
\end{equation}
where
\begin{equation}
T_{ab}=\stackrel{\phi}{T_{ab}}+\stackrel{m}{T_{ab}}\, .
\end{equation}
Assuming that the entropy  is of the form
\begin{equation}
S=\alpha F(\phi) A\, ,
\end{equation}
where $\alpha$ is a constant depending on the number and nature of
quantum fields,  we have
\begin{equation}
\delta S=\alpha \int (\theta F+\dot{F})d\lambda d^{n-1}A\, ,
\end{equation}
where $\theta$ is the expansion of null geodesic congruence and the
overdot means the derivative with respect to the affine parameter
$\lambda$ (do not confuse with the derivative with respect to $t$ in
the previous sections). To extract the $O(\lambda)$ term from the
integrand, we differentiate it with respect to $\lambda$ and use the
requirement ~\cite{Jac1}: $(\theta F+\dot{F})(p)=0$, at the point
$p$, we have
\begin{equation}
\label{diffintegrand} \frac{d}{d\lambda}(\theta
F+\dot{F})|_{\lambda=0}=\dot{\theta}F-F^{-1}\dot{F}^2+\ddot{F}\, .
\end{equation}
By using Raychaudhuri equation for the null geodesic congruence in
$(n+1)$-dimensions
\begin{equation}
\frac{d}{d\lambda}\theta=-\frac{1}{n-1}\theta^2-\sigma_{ab}\sigma^{ab}-R_{ab}k^ak^b\,
\end{equation}
and the geodesic equation, we can rewrite
equation~(\ref{diffintegrand}) as
\begin{equation}
-k^ak^b(R_{ab}-\nabla_a\nabla_bF+F^{-1}\nabla_a
F\nabla_bF)-\frac{1}{n-1}F\theta^2-F\sigma_{ab}\sigma^{ab}\, .
\end{equation}
 We assume that the shear term vanishes in the whole space-time.
With the Clausius relation, we have
\begin{equation}
\label{prestequation} FR_{ab}-\nabla_a\nabla_b
F+\frac{n}{n-1}F^{-1}\nabla_a F\nabla_b F +\Phi g_{ab}
=\frac{2\pi}{\alpha\hbar}(\stackrel{\phi}{T_{ab}}+\stackrel{m}{T_{ab}})\,
.
\end{equation}
where $\Phi$ is an undetermined function and
$\stackrel{\phi}{T_{ab}} = \partial_a\phi\partial_b\phi
-g_{ab}\left(\frac{1}{2}g^{cd}\partial_c\phi\partial_d\phi+V(\phi)\right)\,
$. We assume that the matter stress tensor is divergence free, i.e.,
$\nabla^{a}\stackrel{m}{T_{ab}}=0$, but due to the coupling function
$F(\phi)$ between $\phi$ and scalar curvature $R$, the
$\stackrel{\phi}{T_{ab}}$ is not conserved, namely,
$\nabla^{a}\stackrel{\phi}{T_{ab}} \ne 0$.   Instead the divergence
of $\stackrel{\phi}{T_{ab}}$ is
\begin{equation}
\nabla^{a}\stackrel{\phi}{T_{ab}}=\nabla^2 \phi
\nabla_{b}\phi-V^{'}(\phi)\nabla_b\phi\, .
\end{equation}
Considering the identity
\begin{equation}
\nabla^{a}(FR_{ab}-\nabla_a\nabla_b
F)=\frac{1}{2}\nabla_{b}(FR)-\nabla_b\nabla^2 F-\frac{1}{2}R
F^{'}\nabla_b \phi
\end{equation}
and taking divergence for both sides of the
equation~(\ref{prestequation}), we find
\begin{eqnarray}
\nabla_b \Phi &=&\nabla_b\left(\nabla^2 F-\frac{1}{2}FR\right)
-\nabla^a\left(\frac{n}{n-1}F^{-1}\nabla_a F\nabla_b F\right)\nonumber \\
&&+\left(\frac{1}{2}F^{'}R+\frac{2\pi}{\alpha
\hbar}\left(\nabla^2\phi-V^{'}\right)\right)\nabla_b \phi\, .
\end{eqnarray}
The last term in the right hand side of the above equation is
nothing, but the equation of motion for the scalar field. Therefore
this term always vanishes.  While the left hand side is a pure
gradient of a scalar,  the second term in the right hand side of the
equation $\nabla^a\left(\frac{n}{n-1}F^{-1}\nabla_a F\nabla_b
F\right) $ can not always be written as the gradient of a scalar. As
a result, as the case of $f(R)$ gravity~\cite{Jac1}, there is a
contradiction here. To resolve this, we add an entropy production
term $ d_iS$ to the Clausius relation. It is easy to find if we
choose
\begin{equation}
\label{entropyproduction2} d_i S=\int \sigma d\lambda d^{n-1}A
\end{equation}
with entropy production density
\begin{equation}
\sigma=-\frac{n}{n-1}\alpha F^{-1}\dot{F}^2\lambda
\end{equation}
and use the equation of motion for the scalar field
\begin{equation}
\label{scalarequ} \frac{1}{2}F^{'}R+\frac{2\pi}{\alpha
\hbar}\left(\nabla^2\phi-V^{'}\right)=0\, ,
\end{equation}
 The field equations~(\ref{prestequation}) for the scalar-tensor theory
 become
\begin{equation}
\label{STequ} FG_{ab}-\nabla_a\nabla_bF+g_{ab}\nabla^2
F=\frac{2\pi}{\alpha
\hbar}(\stackrel{\phi}{T_{ab}}+\stackrel{m}{T_{ab}})\, ,
\end{equation}
where $G_{ab}$ is the Einstein tensor.  Taking $\alpha=\frac{1}{4G
\hbar}$, equations~(\ref{STequ}) and~(\ref{scalarequ}) are just the
equations of motion for $g_{ab}$ and $\phi$ in the scalar-tensor
gravity theory, whose Lagrangian is given by (\ref{scalar-tensor}).

Thus we conclude that in order to get the equations of motion in the
scalar-tensor gravity, the Clausius relation has to be modified; we
have to add an entropy production term $d_i S$
(\ref{entropyproduction2}) to the Clausius relation. This suggests
that as the case of $f(R)$ gravity, the scalar-tensor gravity
corresponds to non-equilibrium thermodynamics of space-time. We note
that this entropy production term~(\ref{entropyproduction2}) does
not coincide with the entropy production term~(\ref{eqd_iS}) in the
setup of the FRW universe although they look similar to each other.
This is not surprised because the analysis of Eling {\it et al.} is
locally at each space-time point, while our previous analysis is
focused on the trapping horizon/apparent horizon of the FRW
universe. The common point between them is that both of them imply
that the scalar-tensor gravity requires a non-equilibrium
thermodynamic treatment as the $f(R)$ gravity.

\section{Conclusions}

In this paper we have revisited the relation between the Friedmann
equations and the thermodynamics on the trapping horizon/apparent
horizon in the FRW universe. We have generalized the unified first
law to the case of $(n+1)$-dimensional Einstein theory. After
projecting the unified first law along ``inner" trapping
horizon/apparent horizon, we have obtained the first law of
thermodynamics of the FRW universe, which is very similar to the
thermodynamics of dynamic black holes on their ``outer" trapping
horizon. The form of the first law of thermodynamics is rigorous
without any approximation. For non-Einstein gravity theories, we
have rewritten the field equations to a form of Einstein gravity by
introducing an effective energy-momentum tensor and treated them as
Einstein gravity theory. In these theories the first law of
thermodynamics for the apparent horizon in the FRW universe has the
same form
$$
\langle d E, \xi\rangle=\frac{\kappa}{8\pi G}\langle d A,
\xi\rangle+\langle WdV, \xi\rangle\, .
$$
where $E$ is the Misner-Sharp energy, as the case of Einstein
gravity theory. But here $W$ is an effective work term, the
Misner-Sharp energy is fixed to be the form of
equation~(\ref{MisnerSharp}) because it is defined through
space-time geometry.  In the Lovelock gravity, if we define the heat
$\delta Q$ by pure matter energy-supply projecting along the
horizon, we find that the Clausius relation
$$
\delta Q=Td S
$$
still holds, where $T=\kappa/2\pi$ and $S$ is of the exact form of
the entropy of black hole horizon in the Lovelock gravity. However,
the same treatment tells us that the Clausius relation can not be
fulfilled for the scalar-tensor gravity. We have to introduce an
entropy production term to the Clausius relation
$$
\delta Q=T d S+T d_i S\, .
$$
This implies that the thermodynamics of apparent horizon is a
non-equilibrium thermodynamics for the scalar-tensor theory.

Following Eling {\it et al.}~\cite{Jac1}, we have treated
 the scalar-tensor gravity and shown that the Clausius relation
plus the entropy form $S=\alpha F(\phi) A$ can not give the correct
equations of motion for the theory. In order to resolve this issue,
we have to modify the Clausius relation by introducing an entropy
production term. This also indicates that the scalar-tensor gravity
is the non-equilibrium thermodynamics of spacetime as the case of
$f(R)$ gravity.

\section*{Acknowledgements}
We thank M. Akbar for useful discussions. This research was finished
during R.G.C's visit to the ICTS at USTC, Hefei and the department
of physics of Fudan university, Shanghai, the warm hospitality in
both places extended to him is appreciated.  This work is supported
by grants from NSFC, China (Nos. 10325525 and 90403029), and a
grants from the Chinese Academy of Science.

\end{document}